\documentclass[prd,aps,floats,floatfix,superscriptaddress,preprintnumbers,
showpacs,eqsecnum,nofootinbib,twocolumn]{revtex4}
\usepackage{latexsym,array,theorem,mathrsfs,bm,float}

\usepackage{psfrag}
\usepackage{amsfonts,amsmath,amssymb,latexsym,array,afterpage,theorem,mathrsfs,bm,float,epsfig,color,graphicx,tabularx,here,multirow}
\newcommand{\nn}{\nonumber \\}
\newcommand{\bea}{\begin{eqnarray}}
\newcommand{\ena}{\end{eqnarray}}
\newcommand{\beann}{\begin{eqnarray*}}
\newcommand{\enann}{\end{eqnarray*}}

\newcommand{\ma}[1]{\mbox{$\mathcal{#1}$}}

\newcommand{\calhR}[1]{\raisebox{2ex}{\tiny ({\em h})}\hspace{-0.8em}{\ma R}}

\newcommand{\mpl}{M_{\mathrm{PL}}}

\begin{document}

\title{
Cosmological Dynamics and Double Screening of DBI-Galileon Gravity
}


\author{Sirachak {\sc Panpanich}}
\email{sirachakp-at-gmail.com}
\address{High Energy Physics Theory Group, Department of Physics, 
Faculty of Science, Chulalongkorn University, Phayathai Rd., 
Bangkok 10330, Thailand}
\author{Supakchai {\sc Ponglertsakul}}
\email{supakchai.p-at-gmail.com}
\address{High Energy Physics Theory Group, Department of Physics, 
Faculty of Science, Chulalongkorn University, Phayathai Rd., 
Bangkok 10330, Thailand}
\author{Kei-ichi {\sc Maeda}}
\email{maeda-at-waseda.jp}
\address{Department of Physics, Waseda University, 
Okubo 3-4-1, Shinjuku, Tokyo 169-8555, Japan}


\date{\today}

\begin{abstract}
We investigate cosmological dynamics and screening mechanism of the Dirac-Born-Infeld (DBI) Galileon model.  The model has been divided into two regimes, one has positive signs in front of scalar field kinetic terms so-called the DBI galileon, another one has negative signs and it is dubbed as the DBIonic galileon. We find de Sitter solution and evolution of the Universe starting from radiation dominated era to late-time accelerated expansion in the DBI galileon model without the presence of potential term. In one of the attractors, 
 the ghost and Laplacian instabilities vanishes for the whole evolution. We find mixing of screening mechanisms between the Vainshtein mechanism and the DBIonic  screening mechanism in the DBIonic galileon model, in which a scale changing between these two mechanisms depends on a mass of a source.
\end{abstract}


\maketitle

\section{Introduction}

A number of modified gravity theories have been proposed in order to explain an accelerated expansion of the Universe \cite{Riess:1998cb,Perlmutter:1998np}. Various modified gravity models require extra degree of freedoms, e.g. a scalar field in the Horndeski theories \cite{Horndeski:1974wa,Deffayet:2011gz,Kobayashi:2011nu}, a massive vector field in generalized Proca theories \cite{Heisenberg:2014rta,DeFelice:2016yws}, and a massive graviton field in de Rham-Gabadadze-Tolley (dRGT) massive gravity \cite{deRham:2010kj,deRham:2010ik}. However, after the discovery of GW170817, the higher order terms of the Horndeski and beyond Horndeski theories have been highly constrained  because the quartic and quintic order generally give the sound speed square of the tensor perturbations deviates from unity \cite{Baker:2017hug,Kase:2018iwp,Sakstein:2017xjx} (similar predictions have been made in \cite{Lombriser:2015sxa,Lombriser:2016yzn}). The viable Horndeski theories is now allowed up to cubic order with conformal coupling to gravity. Example models in this class are galileon gravity (up to cubic order) \cite{Nicolis:2008in,Deffayet:2009wt}, kinetic gravity braiding (KGB) model \cite{Deffayet:2010qz}, and $f(R)$ gravity \cite{DeFelice:2010aj}. 

One of the interesting models which is subclass of the cubic Horndeski is the DBI galileon model \cite{deRham:2010eu}. According to \cite{deRham:2016ged} there are three different forms of the generalized galileon Lagrangian which can avoid the caustics singularities, these are pure galileon, DBI galileon, and cuscuta galileon (generalization of the cuscuton \cite{Afshordi:2006ad,Afshordi:2007yx,Afshordi:2009tt} which can be considered as extreme-relativistic limit of the DBI galileon). Cosmology of the pure galileon has already been studied in \cite{DeFelice:2010pv,DeFelice:2010as}. The potential term of scalar field is used to study late-time acceleration and cosmology in the DBI galileon \cite{Sampurnanand:2012rna,Zumalacarregui:2012us}. 
However, the DBI galileon which solely depends on the derivatives of scalar field is also interesting since it satisfies the shift symmetry, $\phi \rightarrow \phi + c$. In addition to the accelerated expansion, screening mechanism is also an essential part of cosmological study. This is because the effect of an extra degree of freedom (a scalar field in the case of DBI galileon) has never been observed in our solar system \cite{Bertotti:2003rm,Williams:2004qba,Kapner:2006si}. Therefore, some screening mechanism is required.

Thus, in this work we will study cosmological dynamics and screening mechanism of the DBI galileon model without the potential term of the scalar field. Action and basic equations of the model are shown in Section  \ref{basiceq}. In Section \ref{desitter} we use dynamical system approach to find de Sitter solution and its stability. In Section. \ref{numer} the de Sitter solution and stabilities are investigated by numerical simulations, and the evolution of density parameters and the equation of state parameters from radiation dominated era to late-time accelerated expansion are discussed in this section. In Section \ref{perturbations} we discuss the ghost and Laplacian instabilities conditions of the model. The screening mechanism has been discussed in the Sec. \ref{screening}, where we obtain an exact solution of spherically symmetric scalar field in the DBIonic galileon. We also investigate the screening behaviour by comparing magnitude of the fifth force with the Newtonian force in this section. Section \ref{conclusions} devotes to our conclusions.


\section{Action and Basic equations}
\label{basiceq}

We consider the cubic Horndeski action with conformal coupling between scalar field and matter field
\bea
S &=& \int d^4 x \sqrt{-g}\left[\frac{1}{2}\mpl^2 R + K(\phi,X) - G_3 (\phi,X) \square \phi \right] \nn
& & + S_m (A^2 (\phi) g_{\mu\nu}, \psi_m)\,, 
\label{action}
\ena
where $R$ is the Ricci scalar, $\mpl$ is the reduced Planck mass, and $A(\phi)$ is a conformal factor. The DBI galileon Lagrangian is obtained by choosing \cite{deRham:2016ged}
\bea
K(\phi,X) &=& a_2 (1 + 2\lambda X)^{1/2} \,,
\label{def_K} \\
G_3(\phi,X) &=& - a_3 \ln (1 + 2\lambda X)  \,,
\label{def_G3} \ena
where $a_i$ and $\lambda$ are constants which can be positive or negative value, and $X = - \frac{1}{2} (\nabla \phi)^2$. Note that we call the theory with positive $\lambda$ as the DBIonic galileon (the DBIonic is a special case of the DBI action in which all the signs of the scalar field Lagrangian have been flipped \cite{Panpanich:2017nft,Burrage:2014uwa}), whereas for negative $\lambda$ it is called the DBI galileon. With the DBI galileon Lagrangian given above, hence we have $K(\phi,X) = K(X)$, $G_3 (\phi,X) = G_3 (X)$.

Varying the action (\ref{action}) with respect to $g^{\mu\nu}$ we find
\begin{equation}
R_{\mu\nu} - \frac{1}{2} g_{\mu\nu} R = \frac{1}{\mpl^2} \left(T_{\mu\nu}^{(\phi)} + T_{\mu\nu}^{(m)} \right) \,,
\end{equation}
where
\bea
T_{\mu\nu}^{(\phi)} &=& \left( K g_{\mu\nu} + K_X \phi_{\mu} \phi_{\nu}\right) - (G_{3X} \square \phi \phi_{\mu} \phi_{\nu} \nn
& & + G_{3;\mu} \phi_{\nu} + G_{3;\nu} \phi_{\mu} - g_{\mu\nu} G_{3;\rho} \phi^{\rho}) \,,
\ena
and $T^{(m)}_{\mu\nu}$ is an energy-momentum tensor of non-relativistic matter and radiation which behave as perfect fluid.
We use short-hand notation as $\phi_{\mu} \equiv \nabla_{\mu} \phi$, $\phi_{\mu\nu} \equiv \nabla_{\mu}\nabla_{\nu} \phi$, and $K_X \equiv \partial K / \partial X$. 

Using the flat Friedmann-Lema\^{\i}tre-Robertson-Walker (FLRW) metric, $ds^2 = - dt^2 + a^2 (t)d \bf{x}^2$, and assuming $\phi = \phi(t)$, we obtain the Friedmann equations as follows
\bea
3\mpl^2 H^2 &=& \rho_{\phi} + \rho_m + \rho_r \,, \label{fdm1} \\
2\mpl^2 \dot H + 3\mpl^2 H^2&=& - (P_{\phi} + P_m + P_r) \,, \label{fdm2}
\ena
where $H\equiv\dot{a}/a$ is the Hubble parameter. We denote derivative with respect to $t$ with ``~$\cdot$~''. $\rho_m$, $\rho_r$, $P_m$, $P_r$ are energy densities and pressures of (non-relativistic) matter and radiation, respectively. The energy density and pressure of the scalar field are given by
\bea
\rho_{\phi} &=& 2X K_X - K + 6H \dot \phi X G_{3X} \,, \\
P_{\phi} &=& K - 2X \ddot \phi G_{3X} \,.
\ena
Variation with respect to $\phi$ leads to the equation of motion of the scalar field
\bea
-2X \ddot \phi K_{XX} - (\ddot \phi + 3H \dot \phi) K_X  + G_{3XX}(-6HX \dot \phi \ddot \phi) \nn 
+ G_{3X} (-6H \dot \phi \ddot \phi - 18 H^2 X - 6\dot H X) = - \frac{A_{,\phi}}{A} T^{(m)} \,. \nn \label{eomscalarfield}
\ena
In this work we choose the conformal factor $A(\phi)$
 as an exponential form with coupling constant $g$ (this should not be confused with the metric determinant)
\bea
A(\phi) = e^{g\phi/\mpl} \,,
\ena
and we assume that the non-relativistic matter is pressureless $P_{m}=0$,
which gives $T^{(m)} \approx - \rho_m$.

In order to obtain dimension of the Lagrangian correctly, we find that
the constants $a_2$, $a_3$ as well as $\lambda$ must have the following dimensions
\bea
\lambda\sim \Lambda^{-4}\,,~
a_2 \sim  \Lambda^{4}\,,~ a_3\sim \Lambda
\,,
\ena
where $\Lambda$ is a constant with mass dimension. 
By use of the scaling of a scalar field $\phi$, we can choose an arbitrary value of the magnitude of $\lambda$ without loss of generality.


\section{de Sitter attractor}
\label{desitter}

\subsection{Dynamical System}

\allowdisplaybreaks
We will study the DBI galileon model, which action is given by Eq. (\ref{action}) with 
(\ref{def_K}) and (\ref{def_G3}).
We find the Friedmann equations (\ref{fdm1})  and (\ref{fdm2}),
in which 
 the energy density and pressure of the scalar field are given by
\beann
\rho_\phi&=& - \frac{a_2}{\sqrt{1+ \lambda \dot \phi^2}} - 6a_3\lambda  \frac{H\dot\phi^3}{1+ \lambda \dot\phi^2} \,,
\\
P_\phi&=& a_2\sqrt{1+ \lambda \dot\phi^2} +2a_3\lambda  \frac{\dot\phi^2 \ddot \phi }{1+\lambda \dot\phi^2} \,.
\enann
The equation of motion of scalar field (\ref{eomscalarfield}) is now 
\bea
& &-  a_2 \lambda\frac{\ddot \phi }{(1 + \lambda \dot\phi^2)^{3/2}} - a_2 \lambda\frac{3H \dot\phi }{\sqrt{1 + \lambda \dot\phi^2}} + 2a_3 \lambda \frac{6H \dot\phi \ddot\phi }{(1+\lambda \dot\phi^2)^2} \nonumber  \\
& & + \frac{2a_3 \lambda}{1+\lambda \dot\phi^2} \left(9H^2 \dot\phi^2 + 3 \dot H \dot\phi^2\right) = \frac{g}{\mpl} \rho_m \,.
\label{eomscalarfield1}
\ena
The energy-momentum conservation gives three equations: 
the radiation energy is conserved as 
\bea
\dot\rho_{r} + 4H \rho_{r} &=& 0 \,,
\label{cons_rad}
\ena
while the conservation of energy-momentum tensor of matter field and scalar field gives 
\beann
\dot\rho_{\phi} + 3H (\rho_{\phi} + P_{\phi}) &=& - \frac{g}{\mpl} \rho_m \dot\phi \,, \\
\dot\rho_{m} + 3H \rho_{m} &=& \frac{g}{\mpl} \rho_m \dot\phi \,, 
\enann
where each energy is not conserved because of the interaction through conformal coupling.

In this paper we use the dynamical system approach to find de Sitter solution and 
analyse its stability. 
We then introduce the dimensionless dynamical variables as follows:
\bea
x_1 \equiv {\frac{\dot\phi}{H_0 \mpl}} \,, ~~ x_2 \equiv {\frac{H}{H_0}} \,, ~~ x_3 \equiv \frac{\rho_r}{3H^2 \mpl^2}=\Omega_r
 \,,\label{dynamicalparameters}
\ena
where $H_0$ is the Hubble parameter at an appropriate time.
For example, we may regard $H_0$ as the present 
Hubble constant, although any value can be chosen.

Eq. (\ref{fdm1}) is rewritten as
\beann
3 \mpl^2 x_2^2
&=& - \frac{a_2}{H_0^2 \sqrt{1 + \tilde \lambda x_1^2}} 
-  \frac{6a_3 \tilde \lambda  \mpl x_2x_1^3 }{1 + \tilde \lambda x_1^2} + {\frac{\rho_m}{H_0^2}} \nn
& & + 3 x_2^2 x_3 \mpl^2 \,,
\enann
where 
$\tilde \lambda=\lambda H_0^2 \mpl^2$.
The energy density and pressure of the scalar field are given by
\beann
{\frac{\rho_\phi}{H_0^2}}&=& - \frac{a_2}{H_0^2 \sqrt{1+ \tilde \lambda x_1^2}} -  \frac{ 6a_3 
\tilde \lambda \mpl x_2 x_1^3}{1+ \tilde \lambda x_1^2} \,,
\\
{\frac{P_\phi}{H_0^2}}&=& \frac{a_2}{H_0^2} \sqrt{1+ \tilde \lambda x_1^2} +
 \frac{2a_3 \tilde \lambda\mpl x_1^2}{1+\tilde \lambda x_1^2}
\times{\frac{\ddot \phi}{H_0^2\mpl}} \,.
\enann
By use of the freedom of $\lambda$, we 
 fix  $\tilde \lambda =-1$ or $1$, i.e., 
 $\lambda=1/H_0^2 \mpl^2$.
We also assume
\beann
\frac{a_2}{H_0^2\mpl^2} =\frac{\alpha_2}{ \tilde \lambda}
\,,~~{\rm and}~~~
\frac{a_3}{\mpl} =\frac{\alpha_3}{ 2\tilde \lambda } 
\,,
\enann
where $\alpha_2$ and $\alpha_3$ are positive dimensionless constants. For simplicity, we set $\alpha_2 = \alpha_3 = 1$.
In what follows, we drop the tilde for brevity.
Note that we are interested in the expanding Universe $(H>0)$, which means $x_2>0$.
Since the  radiation energy is positive, we impose $x_3\geq 0$.
However, $x_1$ can be negative or positive depending on dynamics of the scalar field.

With these dynamical variables, the Friedmann equation (\ref{fdm1}) is written as
\bea
1 = - \frac{1}{3x_2^2 \lambda \sqrt{1 + \lambda x_1^2}} - \frac{x_1^3}{x_2 (1 + \lambda x_1^2)} + \Omega_m + x_3 \,,
\ena
which is regarded as a constraint equation since it contains no derivative term. 
In addition, the density parameters of matter, scalar field, and radiation are defined respectively as 
\bea
\Omega_m &=& 1 +  \frac{1}{3x_2^2 \lambda \sqrt{1 + \lambda x_1^2}} + \frac{x_1^3}{x_2 (1 + \lambda x_1^2)} - x_3 \,,
 \label{omegam}
\\
\Omega_{\phi} &=&  - \frac{1}{3x_2^2 \lambda \sqrt{1 + \lambda x_1^2}} - \frac{x_1^3}{x_2 (1 + \lambda x_1^2)} \,,
 \label{omegaphi}
\\
\Omega_r &=& x_3 
\label{omegar}\,.
\ena
Differentiating the dynamical parameters with respect to the e-folding number ($N = \ln a$), 
we find 
\bea
\frac{dx_1}{dN} &=& \frac{\ddot \phi}{H_0^2\mpl x_2} \,, \label{autono1}\\
\frac{dx_2}{dN} &=& \frac{\dot H}{H_0^2x_2} \,, \label{autono2} \\
\frac{dx_3}{dN} &=& -2\frac{x_3}{x_2^2}\left({\frac{\dot H}{H_0^2}}+2x_2^2\right) \,. \label{autono3}
\ena
In the last equation, we have used the conservation equation of radiation energy
(\ref{cons_rad}).
\begin{widetext}
From the second Friedmann equation (\ref{fdm2}) and the equation of motion for the scalar field (\ref{eomscalarfield1}), we find
\bea
{\frac{\ddot \phi}{H_0^2\mpl}}&=& -\frac{\Big(\lambda  x_1^2+1\Big)}{\Delta} \Big\{2 g \Big[
3\lambda x_2\Big(x_1^3-(\lambda x_1^2+1)x_2 (x_3-1)\Big)+\sqrt{\lambda x_1^2+1}\Big]
 \nn
& &
+3 \lambda x_1^2 x_2^2(x_3-3)+3x_1(x_1+2 \lambda x_2)\sqrt{\lambda x_1^2+1}\Big\} \,,
\label{eq_ddotphi}
\\
{\frac{\dot H}{H_0^2}}
&=& \frac{1}{\Delta}\left\{g x_1^2
\left[3 \lambda x_2 \left(
x_1^3 - (\lambda x_1^2+1) x_2 (x_3-1)\right)
+ \sqrt{\lambda  x_1^2+1}\right] \right.
 \left.+
 3\lambda x_1 x_2^2\left[-3   x_1^3 +2  x_2  (x_3+3)\right]
 -(\lambda  x_1^2+1)
  \right.
\nn
& &
\left.
+x_2 \left[6 x_1  -\lambda  x_2 x_3  
+3  \lambda ( x_1^3 -x_2)\right]\sqrt{\lambda  x_1^2+1}
\right\} \,, 
\label{eq_dotH}
\ena
where
\beann
\Delta &\equiv& \lambda  \left(3x_1(x_1^3-4x_2)+2 \sqrt{\lambda  x_1^2+1}\right) \,.
\hskip 7cm
\enann
Substituting Eqs. (\ref{eq_ddotphi}) and (\ref{eq_dotH}) into Eqs. (\ref{autono1}),  (\ref{autono2})  and  (\ref{autono3}) , 
we find  the autonomous equations as
\bea
\frac{d x_1}{dN} &=& -\frac{\left(\lambda  x_1^2+1\right) }{\Delta x_2} \Big\{2 g \Big[
3\lambda x_2\Big(x_1^3-(\lambda x_1^2+1)x_2 (x_3-1)\Big)+\sqrt{\lambda x_1^2+1}\Big]
 \nn
& &
+3\lambda x_1^2 x_2^2(x_3-3)+3x_1(x_1+2\lambda x_2)\sqrt{\lambda x_1^2+1}\Big\}\,, \label{autoeq1}\\
\frac{d x_2}{dN}&=&  \frac{1}{\Delta x_2}\left\{g x_1^2
\left[3 \lambda x_2 \left(
x_1^3- (\lambda x_1^2+1) x_2 (x_3-1)\right)
+ \sqrt{\lambda  x_1^2+1}\right] \right.
   \left.
 +
 3\lambda x_1 x_2^2\left[-3 x_1^3 +2  x_2  (x_3+3)\right]
 -(\lambda  x_1^2+1)
  \right.
\nn
& &
\left.
+x_2\left[6 x_1  -\lambda  x_2 x_3  
+3 \lambda ( x_1^3 -x_2)\right]\sqrt{\lambda  x_1^2+1}
\right\} \,, 
 \label{autoeq2} \\
\frac{dx_3}{dN} &=& -\frac{2 x_3}{\Delta x_2^2} 
 \Big\{g x_1^2 \left[3 \lambda x_2 \left(
x_1^3- (\lambda x_1^2+1) x_2 (x_3-1)\right)
+ \sqrt{\lambda  x_1^2+1}\right] +3\lambda x_1 x_2^2[-x_1^3+2x_2(x_3-1)]-(\lambda x_1^2+1)
\nn
& &
+x_2\left[6x_1-\lambda x_2 x_3+\lambda (3x_1^3+ x_2)
\right]\sqrt{\lambda x_1^2+1}\Big\}
 \,. \label{autoeq3}
\ena
Since $\lambda$ and $g$ are constant, these autonomous equations are closed. 
\end{widetext}

\subsection{Fixed points without matter coupling and their stability ($g=0$)}

\subsubsection{fixed points}
First in the case of no conformal coupling ($g=0$), 
we analyze the fixed points of the autonomous system, 
which are obtained by setting $dx_1/dN = dx_2/dN = dx_3/dN = 0$.
Because of $\dot H = 0$ and $H>0$,  these fixed points give de Sitter spacetime. 
Note that in order to obtain purely de Sitter solution, we require $\Omega_{\phi} = 1$, while $\Omega_m = 0$.

From Eqs. (\ref{autoeq2}) and   (\ref{autoeq3}), we obtain 
\bea
\frac{dx_3}{dN} = - x_3 \left( \frac{2}{x_2} \frac{dx_2}{dN} + 4 \right) \,, \label{relation}
\ena
which is just the conservation of radiation energy. 
It can be easily integrated as
\beann
\ln (x_3 x_2^2)=-4N+{\rm constant}\,,
\enann
which is equivalent to $\rho_r\propto a^{-4}$.

When we are interested in the fixed points, they must be constants in time.
Since we assume the expanding universe, we find that $x_3=0$ at the fixed points,
which means the radiation energy density must vanish asymptotically.

The fixed points are given in Table \ref{fixedpoint}.
It is obvious that the DBIonic galileon ($\lambda = +1$) does not provide real fixed points. In contrast, there are two real fixed points, (a) and (b), in the DBI galileon case ($\lambda = -1$),
which means there exist two de Sitter solutions. 

\begin{table}[h]
\begin{center}
  \begin{tabular}{|c||c|c|c|}
\hline 
& & & \\[-.5em]
$\lambda = -1 $&$x_1$&$x_2$ &$~~ x_3 ~~$
\\[.5em]
\hline
& & & \\[-.5em]
(a)&$0$&$\frac{1}{\sqrt{3}}$&$0$
\\[.5em]
\hline
& & & \\[-.5em]
(b)&$\frac{1}{3} \sqrt{\frac{1}{2} \left(\sqrt{37}-1\right)}$&$\frac{1}{3} \sqrt{\frac{1}{2} \left(\sqrt{37}-1\right)}$&$0$
\\[.5em]
\hline
\hline
& & & \\[-.5em]
$\lambda = +1 $&$x_1$&$x_2$ &$x_3$
\\[.5em]
\hline
& & & \\[-.5em]
(c)&$0$&$\pm \frac{i}{\sqrt{3}}$&$0$
\\[.5em]
\hline
& & & \\[-.5em]
(d)&$\pm\frac{i}{3} \sqrt{\frac{1}{2} \left(\sqrt{37}-1\right)}$&$\mp\frac{i}{3} \sqrt{\frac{1}{2} \left(\sqrt{37}-1\right)}$&$0$
\\[.5em]
\hline 

 \end{tabular}
    \caption{The fixed points of the autonomous equations (\ref{autoeq1}) -  (\ref{autoeq3}) with $g = 0$ where $\lambda = -1$ refers to the DBI galileon case, while $\lambda = +1$ is the DBIonic galileon case.}
\label{fixedpoint}
\end{center}
\end{table}

The fixed point (a) gives $x_1 = 0$, which means that the kinetic term of the scalar field vanishes, and then the energy density of the scalar field becomes a constant,
 which is similar to a cosmological constant model. 
On the other hand, for the fixed point (b), $x_1 \not= 0$, and then the kinetic term still exists.
Thus the fixed point gives different behaviour from the former case.

\begin{widetext}
The equation of state of the scalar field and the effective equation of state are given by
\bea
w_{\phi} &=& \frac{P_{\phi}}{\rho_{\phi}} = \frac{\lambda^{-1} (1 + \lambda x_1^2)^{3/2} + x_1^2 \displaystyle{\left(\frac{\ddot \phi}{H_0^2 \mpl}\right)}}{x_1^2 \sqrt{1+ \lambda x_1^2} - \lambda^{-1} (1 + \lambda x_1^2)^{3/2} - 3x_2 x_1^3} \,,
\nn
&=&
- \frac{(1-x_1^2)\left[6x_1 x_2(2-x_1^2)\sqrt{1-x_1^2}-\left(
2(1-x_1^2)+3x_1^4x_2^2(x_3-3)\right)
\right]}{\left(3 x_1^3 x_2-\sqrt{1-x_1^2}\right) \left(2 \sqrt{1-x_1^2}+3
x_1(x_1^3-4x_2)\right)} \,.
 \label{wphi}
\\
w_{\rm eff} &=& -1 -\frac{2}{3}\frac{\dot H}{H^2} = -1 -\frac{2}{3}\frac{\dot H}{H_0^2 x_2^2} 
\nn
&=&\frac{9 x_1^4 x_2^2+12 x_1 x_2 \left(\sqrt{1-x_1^2}-x_2^2 x_3\right)+2 \sqrt{1-x_1^2} x_2^2 x_3+2 x_1^2-6 \sqrt{1-x_1^2} x_1^3 x_2-2}{3 x_2^2 \left(2 \sqrt{1-x_1^2}+3
x_1(x_1^3-4x_2)\right)}
\,, \label{weff}
\ena
respectively, where we have used Eqs. (\ref{eq_ddotphi}) and (\ref{eq_dotH}) with $g=0$ and $\lambda = -1$.
\\
\end{widetext}

Both fixed points (a) and (b) of the DBI galileon  give the same density parameters and the same equation-of-state parameters
as $ \Omega_\phi=1, \Omega_m=0, \Omega_r=0$ and 
$w_\phi=w_{\rm eff}=-1$, respectively.
 The results confirm that the both fixed points are purely de Sitter solutions.

\subsubsection{Stability}
 
Here we discuss stability of the above fixed points.
We start by rewriting the autonomous equations such that
\beann
\frac{dx_1}{dN} &=& \mathcal{F}(x_1,x_2,x_3) \,, \nn
\frac{dx_2}{dN} &=&  \mathcal{G}(x_1,x_2,x_3) \,, \nn
\frac{dx_3}{dN} &=& \mathcal{H}(x_1,x_2,x_3) \,. 
\enann

Then stability of de Sitter solutions will be investigated by using the linear perturbation
 analysis around each fixed point
$(x_1^{(c)}, x_2^{(c)}, x_3^{(c)})$, i.e., by setting
\beann
x_i(N) &=& x_i^{(c)} + \delta x_i(N) \,, 
\enann
($i=1,2,3$), we find the perturbation equations as
\bea
\frac{d}{dN} \begin{pmatrix} \delta x_1 \\ \delta x_2 \\ \delta x_3 \end{pmatrix} =
 \mathcal{M} \begin{pmatrix} \delta x_1 \\ \delta x_2 \\ \delta x_3 \end{pmatrix} \,, \label{firstordercouple}
\ena
where 
\beann
\mathcal{M} = \left.\begin{pmatrix} \frac{\partial \mathcal{F}}{\partial x_1} && \frac{\partial \mathcal{F}}{\partial x_2} &&  \frac{\partial \mathcal{F}}{\partial x_3} \\[.5em]
\frac{\partial \mathcal{G}}{\partial x_1} && \frac{\partial \mathcal{G}}{\partial x_2} &&  \frac{\partial \mathcal{G}}{\partial x_3} \\[.5em]
\frac{\partial \mathcal{H}}{\partial x_1} && \frac{\partial \mathcal{H}}{\partial x_2} &&  \frac{\partial \mathcal{H}}{\partial x_3}  \end{pmatrix}\right|_{x_1^{(c)},x_2^{(c)},x_3^{(c)}} \,.
\enann
Complete expression of each component of the matrix $\mathcal{M}$ is shown in Appendix \ref{matrixperturbations}.

The first order coupled differential equation (\ref{firstordercouple}) has a general solution as
\beann
\delta x_i \propto e^{\mu N}\,,
\enann
where $\mu$ is an eigenvalue of the matrix $\mathcal{M}$. If all eigenvalues are negative (or their real parts are negative for complex eigenvalues), the fixed point is stable.
 If  one or two eigenvalues (or their real parts for complex ones)  are positive, the fixed point is a saddle point. Lastly, if all of eigenvalues are positive, the fixed point is unstable.

Now we show the results for the DBI galileon ($\lambda=-1$) with $g=0$.
For the fixed point  (a), 
the three eigenvalues are
\bea
\mu&=&~-4\,, ~ -3\,, ~ -3 \,.
\ena
Then the fixed point (a) is stable.
While for the fixed point  (b), 
the corresponding eigenvalues are
\bea
\mu&=&~-4\,, ~ -3\,, ~ -3 \,.
\ena
Thus the fixed point (b) is also stable. 

The difference between the two de Sitter points can be understood by numerical simulations and cosmological perturbations which will be shown in the following sections.

\subsection{Fixed points with matter coupling and their stability ($g\neq 0$)}
\begin{figure}[h]
	\centerline{\includegraphics[width=7cm]{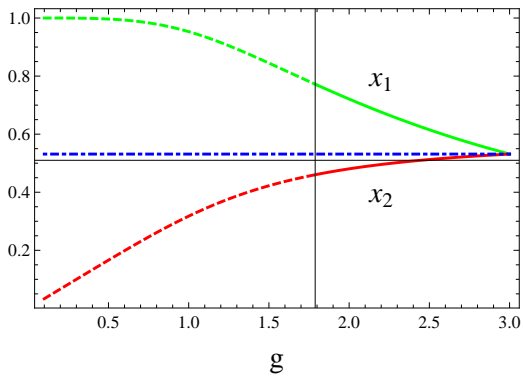}}
	\caption{The fixed point with $g \neq 0$ where thick lines represent in the range $g_{\rm cr} (\sim 1.788) \leq g < 3$ and dashed lines represent in the range $0 < g < g_{\rm cr}$. The fixed point (b) is represented as a blue dotted-dashed line, where $x_3 = 0$.}
\label{x1x2vsg}
\end{figure}

\begin{figure}[h]
	\centerline{\includegraphics[width=5.8cm]{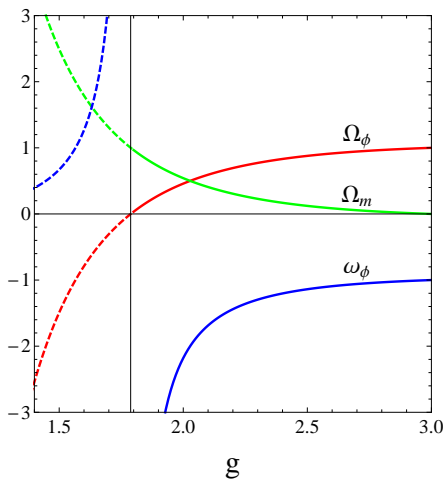}}
	\caption{The density parameters , $\Omega_m$ (green line) and $\Omega_{\phi}$ (red line), and the equation of state of the scalar field (blue line) of the fixed point with $g \neq 0$. The thick lines represent values in the range $g_{\rm cr} (\sim 1.788) \leq g < 3$, while $0 < g < g_{\rm cr}$ is represented as dashed lines.}
\label{omegavsg}
\end{figure}

When there is a coupling between matter field and the scalar field, 
we have the same fixed points (a) and (b) as the previous ones.
We also find another fixed point for the DBI galileon ($\lambda = -1$)
 in the range of coupling constant $0 <g<3$. 
We show the numerical result  in Figs. \ref{x1x2vsg} and \ref{omegavsg}.
In Fig. \ref{x1x2vsg}, 
we show the values of fixed points $x_1$ and $x_2$ in terms of 
the coupling constant $g$.
As $g\rightarrow 3$, both values $x_1$ and $x_2$ approaches to those of the fixed point (b). 
There is no third fixed point beyond $g=3$. 

We also depict the values of the density parameters $\Omega_m$ and $\Omega_\phi$ 
in Fig. \ref{omegavsg}.
From Fig. \ref{omegavsg}, we find that $\Omega_\phi \geq 0$ only when 
 $g_{\rm cr} \leq g<3$, where $g_{\rm cr}\approx 1.788364231$.
 The parameter range $0< g<g_{\rm cr} $ may not be preferable for the cosmological model because 
the energy density of the scalar field becomes negative.
 We also show the equation-of-state parameter $w_\phi$ in Fig. \ref{omegavsg}.
 We find the scalar field behaves just as a phantom matter in the range of  $g_{\rm cr} \leq g<3$
 because $w_\phi<-1$. 
 
Since both density parameters do not vanish, 
 this fixed point corresponds to the so-called scaling solution \cite{Copeland:1997et}. 
  However since $\dot H=0$ at this fixed point, it is also de Sitter spacetime although 
 there exists non-zero constant matter density.
 The energy density of the scalar field is converted into matter density in order to keep 
 it constant.

 \begin{figure}[h]
	\centerline{\includegraphics[width=8.3cm]{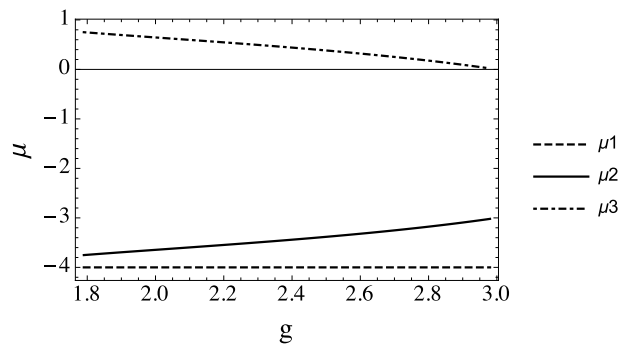}}
	\caption{Eigenvalues of the fixed point with $g \neq 0$ in the range $g_{\rm cr} (\sim 1.788) \leq g < 3$.}
\label{eigenvalues}
\end{figure}
 
It is worth noting that  for  $g = 2.2$, we find the fixed point with $x_1 = 0.675886$, $x_2 = 0.49565$, which yields the present observed values of the density parameters, that is,
$\Omega_m\approx 0.305824$ and $\Omega_\phi\approx 0.694176$.
 
Since the scalar field at this fixed point shows a phantom-like behaviour,
 we should  check its stability.
The result of the stability analysis is depicted in Fig. \ref{eigenvalues}, which reveals that
 this fixed point is a saddle (one eigenvalue $\mu$ is always positive
 for $g_{\rm cr} \leq g<3$. 
Since it is not an attractor, the fine-tuning for the initial values is required to find this fixed point.  Hence in what follows, we will no longer consider this type of fixed points.


\section{Numerical Solution}
\label{numer}

\subsection{Trajectories of solutions}
We have two stable fixed points.
Which solution is realized when the universe evolves?
In order to find the answer, we will study the global stability which analysis requires numerical study. 

\begin{figure}
	\includegraphics[width=8cm]{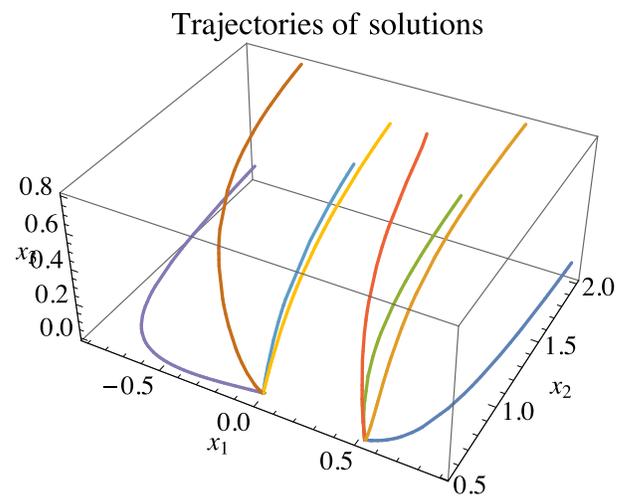}
	\caption{Trajectories of solutions of the autonomous equations (\ref{autoeq1}) - (\ref{autoeq3}) where we set initial conditions as $x_2 = 2$, and allow $x_1$ and $x_3$ to be arbitrary.}
\label{trajectories1}
\end{figure}

We solve Eqs. (\ref{autoeq1}) - (\ref{autoeq3}) numerically with $g=0$ and $\lambda = -1$. Trajectories of solutions are shown in Fig. \ref{trajectories1} where we initially set $x_2 = 2$
 while arbitrarily choose the initial value of $x_1$ (kinetic term of the scalar field) and $x_3$ (density parameter of the radiation). Obviously, for negative $x_1$ the solutions converge to the fixed point (a), whereas for positive $x_1$ the fixed point (b) is an attractor. 

\vskip 1cm

\begin{figure}
	\includegraphics[width=8cm]{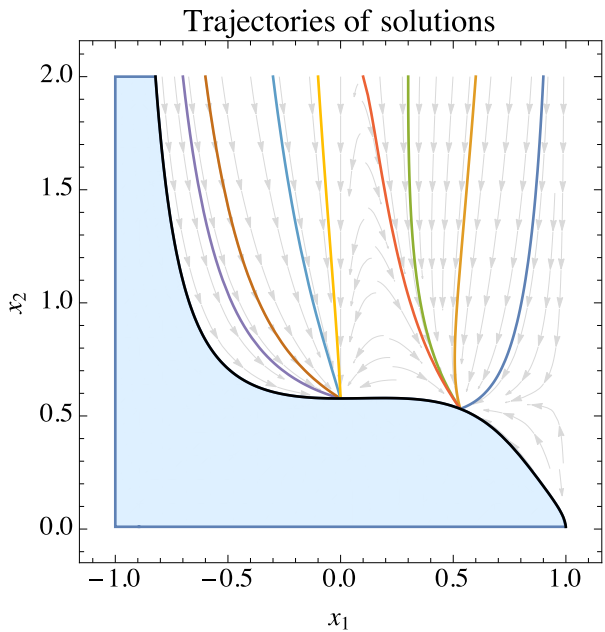}
	\caption{2D Trajectories of solutions of two autonomous equations (\ref{autoeq1}) and (\ref{autoeq2}) where we assume $x_3 = 0$. We set initial conditions as $x_2 = 2$, and let $x_1$ to be arbitrary. The blue region is a forbidden region since it yields $\Omega_{\phi} > 1$, and the black-thick line is $\Omega_{\phi} = 1$. }
\label{trajectories2}
\end{figure}

In order to see behaviour of the solutions more clearly, assuming $x_3 = 0$ 
(it is true if we can initially ignore the radiation density), we show the trajectories 
of the evolution in Fig. \ref{trajectories2}.
We set initial value of $x_2 = 2$.
 The blue region is a forbidden region because $\Omega_{\phi} > 1$. The black-thick line is $\Omega_{\phi} = 1$. Again, this figure confirms that for solutions with $x_1 > 0$ and large $x_2$ the trajectories evolve to the fixed point (b), while $x_1 < 0$ converges to the fixed point (a). 
Consequently, 
although 
both fixed points are stable, which fixed point is realized
 depends on the initial condition of $x_1$.


\subsection{Cosmological Evolution}

\begin{figure}[h]
\includegraphics[width=7.5cm]{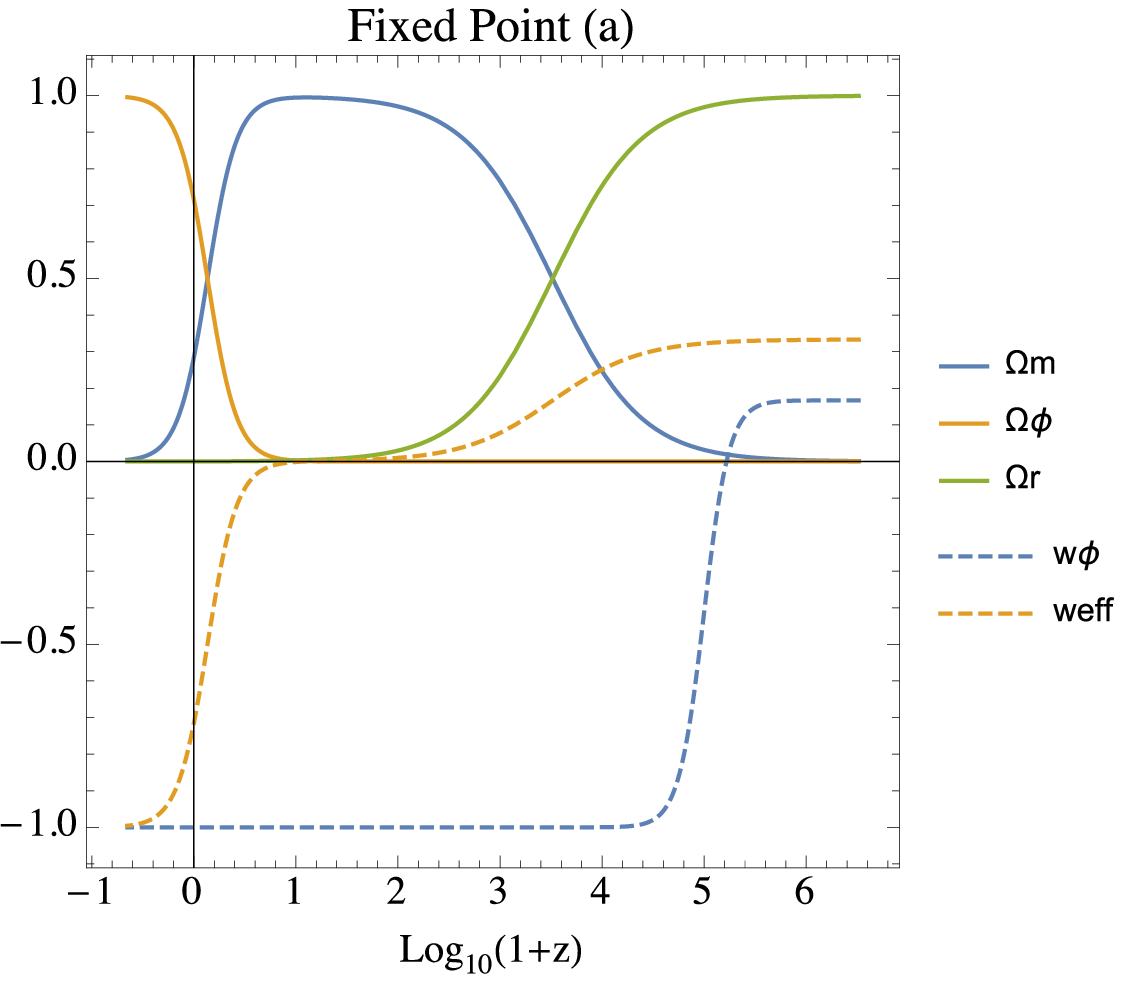}
\\
\includegraphics[width=7.5cm]{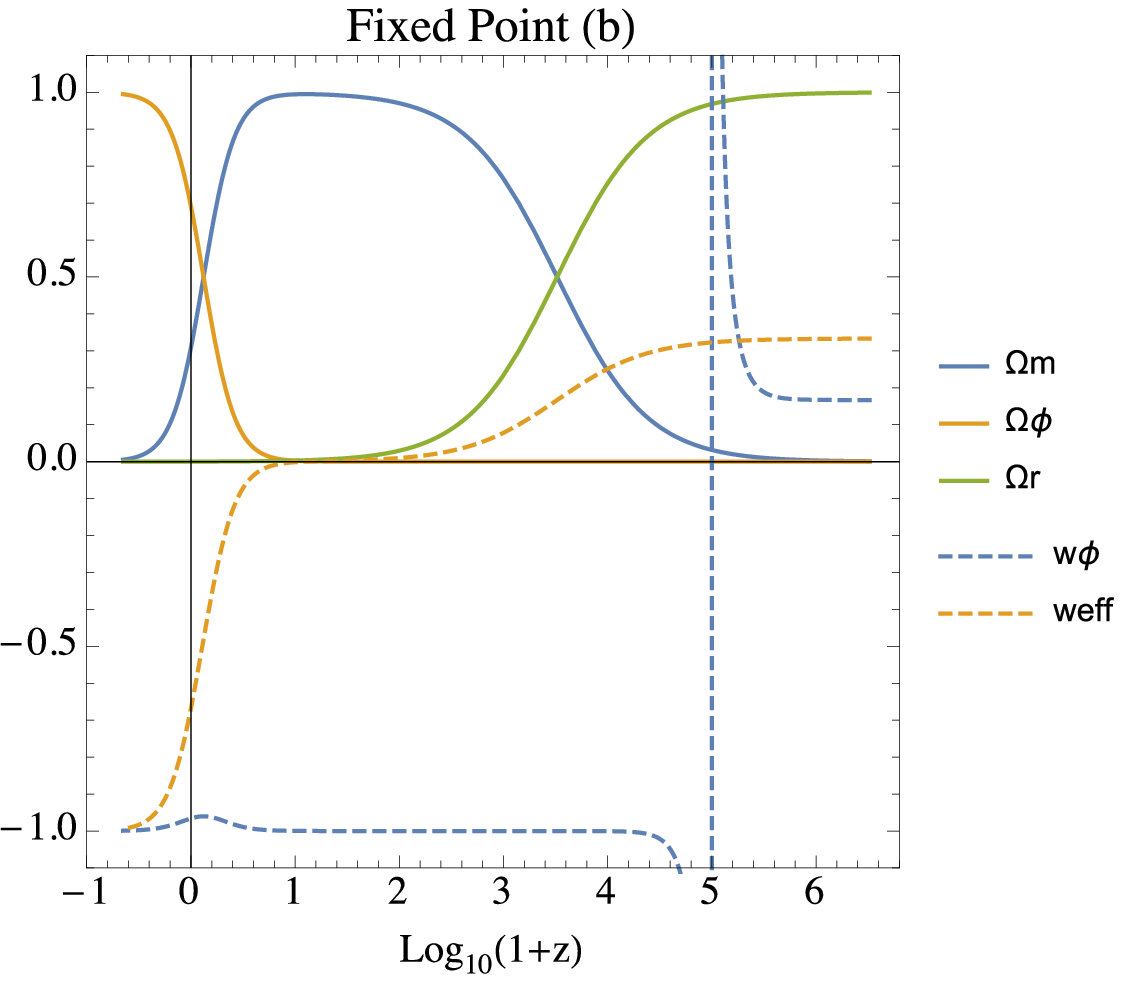}\caption{Evolution of the density parameters and the equation of state parameters of the fixed point (a) (top) where initial conditions are $x_3 = 0.999$, $x_2 = 6.8 \times 10^{10}$, and $x_1 = -0.01$ at $z = 3.27 \times 10^6$. For the fixed point (b) (bottom) we choose initial conditions as $x_3 = 0.999$, $x_2 = 6.8 \times 10^{10}$, and $x_1 = +0.01$ at $z = 3.27 \times 10^6$.}
\label{evolution}
\end{figure}

To answer whether we can obtain a realistic evolution of the universe, 
we solve Eqs. (\ref{autoeq1}) - (\ref{autoeq3}) from $z = 3.27 \times 10^6$ (radiation dominated era) to $z = -0.77$ (de Sitter expansion stage) for the case of DBI galileon model ($\lambda = -1$) with $g=0$. The evolution of density parameters (\ref{omegam}) - (\ref{omegar}), and the equation of state parameters (\ref{wphi}) and (\ref{weff}) 
are shown in Fig. \ref{evolution}.

Both fixed points (a) and (b) provide a successful cosmological evolution, namely, starting from a radiation dominated era to a matter dominated era, and following by a dark energy dominated era or accelerated expansion epoch. The difference between these two fixed points is the evolution of $w_{\phi}$, which is given by 
Eq. (\ref{wphi}).

For $x_1 > 0$ (the fixed point  (b)),  the $w_{\phi}$  diverges  at 
$\log_{10} (1+z) \approx 5$, 
 as shown in Fig. \ref{evolution},
 because the term $\left(3 x_1^3 x_2-\sqrt{1-x_1^2}\right)$ in the denominator vanishes there 
 but other terms are finite.



\section{Ghosts and Laplacian instability} 
\label{perturbations}

\begin{figure}[htb]
	\includegraphics[width=8.5cm]{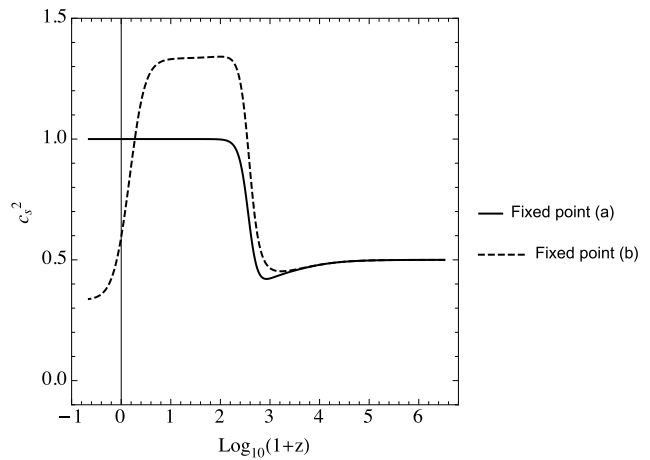}
	\caption{Evolution of the sound speed square of the DBI galileon model where initial conditions are the same as in the Fig. \ref{evolution}.}
\label{soundspeed}
\end{figure}

\begin{figure}[htb]
	\includegraphics[width=6.6cm]{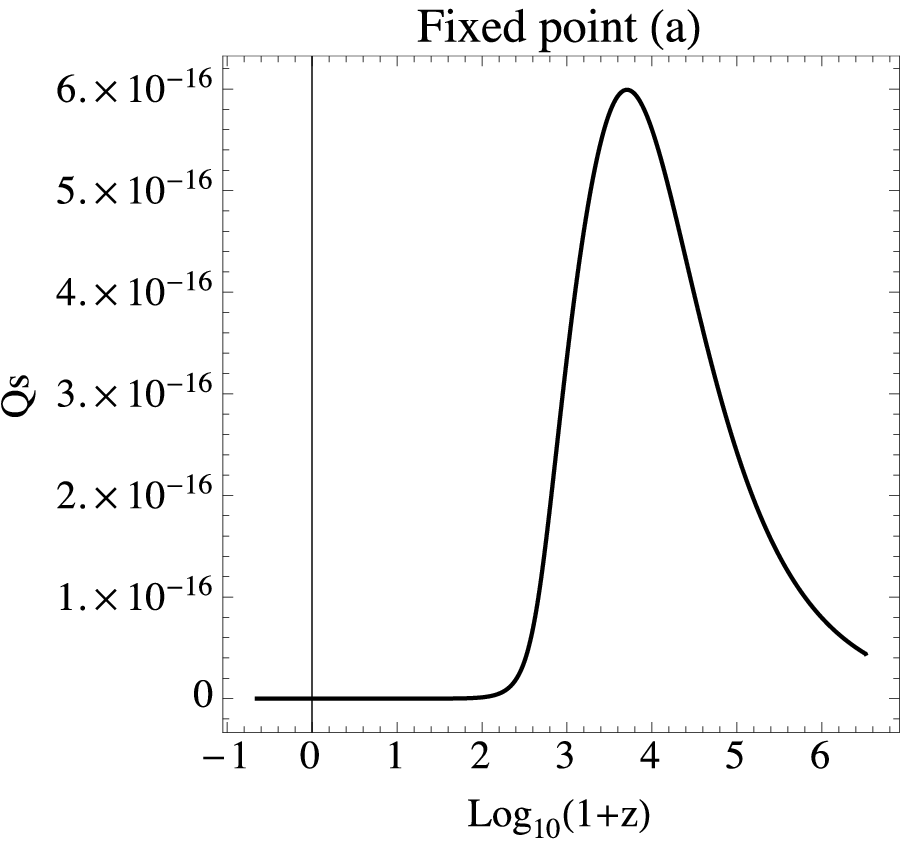}
	\includegraphics[width=6cm]{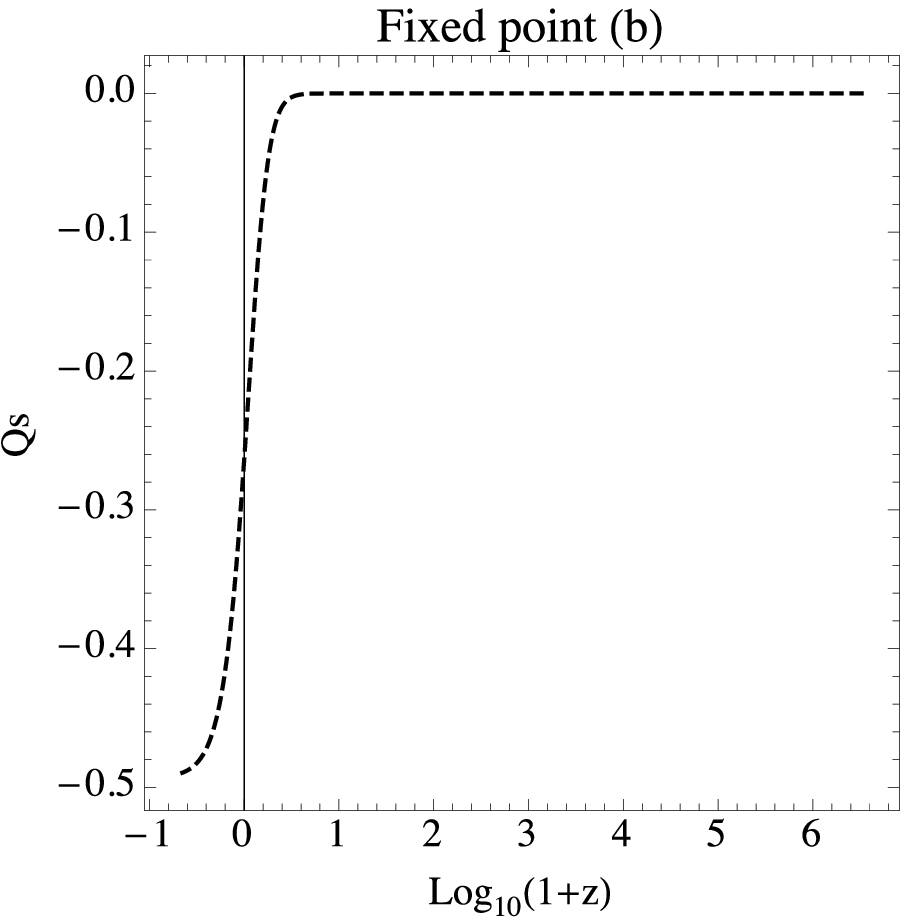}
	\caption{Evolution of the $Q_s$ parameter of the fixed point (a) (top) and the fixed point (b) (bottom). The initial conditions are the same as given in the Fig. \ref{evolution}.}
\label{coefficient}
\end{figure}

Now we analyze whether our evolutional scenario has no ghost or Laplacian instability.
According to \cite{Tsujikawa:2014mba,Gleyzes:2013ooa,Gergely:2014rna,Kase:2014yya}, by using Arnowitt, Deser and Misner (ADM) 3+1 decomposition and choosing unitary gauge, namely a timelike hypersurface coincides with a constant $\phi$ hypersurface, 
we write down the second-order Lagrangian of the Horndeski theory. 
For the scalar perturbations, we expand the Lagrangian density up to the second-order of the 
perturbations around the background flat FLRW metric; 
\bea
ds^2 &=& - e^{2A} dt^2 + 2 \psi_{|i} dx^i dt \nn
& & + a^2 (t) (e^{2\zeta}\delta_{ij} + \partial_{ij} E) dx^i dx^j \,, \label{pertFLRWmetric}
\ena
where $A, \psi, \zeta$, and $E$ are scalar metric perturbations. The $|i$ is a covariant derivative with respect to the spatial metric $h_{ij}$, and $\partial_{ij} = \nabla_{i} \nabla_{j} - \delta_{ij} \nabla^2/3$. We choose the gauge such that $E = 0$, hence $h_{ij} = a^2 (t) e^{2\zeta} \delta_{ij}$. To avoid the Ostrogradsky instability, we require that spatial derivative and mixing of time and spatial derivative term that are higher than second-order must be vanished. Then by substituting this metric (\ref{pertFLRWmetric}) into the second-order Lagrangian density, we find
\bea
S_s^{(2)} = \int d^4 x a^3 Q_s \left[\dot\zeta^2 - \frac{c_s^2}{a^2} (\partial \zeta)^2\right] \,,
\ena
where $\zeta$ is the curvature perturbation and
 the coefficient of the kinetic term and squared sound speed are defined \cite{DeFelice:2011bh}
 by 
\bea
Q_s &\equiv& \frac{w_1(4w_1 w_3 + 9w_2^2)}{3w_2^2} \,,
\ena
\begin{widetext}
\bea
c_s^2 &\equiv& \frac{3(2w_1^2 w_2 H - w_2^2 w_4 + 4 w_1 w_2 \dot w_1 - 2w_1^2 \dot w_2) - 6w_1^2 [(1+w_m)\rho_m + (1+w_r)\rho_r]}{w_1(4w_1 w_3 + 9 w_2^2)} \,. 
\ena
\end{widetext}
Remark that, the perfect fluid part of the above sound speed square equation can be proved from cosmological perturbations of the Horndeski theory with additional k-essence scalar fields \cite{Gergely:2014rna,Kase:2014yya}. 
For the DBI galileon (recall that $\lambda = -1$), parameters $w_i$ are 
given in terms of the dynamical variables  (\ref{dynamicalparameters}) as
\bea
w_1 &=& 1 \,, \\
w_2 &=& \frac{x_1^3}{1 + \lambda x_1^2} + 2x_2 \,, \\
w_3 &=& 3x_1^2 \left(\frac{1}{2\sqrt{1 + \lambda x_1^2}} - \frac{\lambda x_1^2}{2 (1 + \lambda x_1^2)^{3/2}} \right. \nn
& & \left. + \frac{3x_1^3 x_2 \lambda}{(1 + \lambda x_1^2)^2} - \frac{6x_1 x_2}{1 + \lambda x_1^2}\right) - 9x_2^2 \,, \\
w_4 &=& 1 \,.
\ena
To avoid the ghost instability the coefficient in front of the kinetic term $\dot\zeta^2$ must be positive, and the sound speed square of the scalar perturbations also needs to be positive to avoid the Laplacian instability. Thus, we require
\bea
Q_s > 0 \,, ~~ c_s^2 > 0 \,.
\ena
By using the same initial conditions as previously used in cosmological evolution Sec. \ref{numer}, the squared sound speed and the kinetic coefficient of each fixed point are represented in Fig. \ref{soundspeed} and \ref{coefficient}, respectively.

According to Fig. \ref{soundspeed} we find that both fixed points have $c_s^2 > 0$ for the whole evolution, then they satisfy the Laplacian stability condition. Nevertheless, the fixed point (b) has $c_s > 1$ during the matter dominated epoch. Fig. \ref{coefficient} shows that the fixed point (b) has $Q_s < 0$ at late-time, thus only the fixed point (a) can avoid the ghost instability. Therefore, at the perturbations level, only the fixed point (a) is the viable solution.

For the tensor perturbations, the three-dimensional metric is
\beann
h_{ij} = a^2 (t) e^{2\zeta} \hat h_{ij} \,, ~~ \hat h_{ij} = \delta_{ij} + \gamma_{ij} + \frac{1}{2} \gamma_{il}\gamma_{lj} \,.
\enann
where $\det \hat h = 1$ and $\gamma_{ij} = \partial_i \gamma_{ij} = 0$. By inserting this three-dimensional metric into the second-order Lagrangian density and set all of scalar perturbations to zero. We find 
\bea
S_T^{(2)} = \frac{1}{2}\int d^4 x a^3 Q_T\left(\dot \gamma_{ij}^2 - \frac{c_T^2}{a^2} (\partial_k \gamma_{ij})^2\right) \,.
\ena
In the similar way, the conditions for avoidance of the ghost and Laplacian instabilities are
\bea
Q_T > 0 \,, ~~ c_T^2 > 0 \,.
\ena
For the DBI galileon, the kinetic coefficient and the sound speed square of the tensor perturbations are
\bea
Q_T = \frac{w_1}{4} = \frac{1}{4} > 0 \,, \\
c_T^2 = \frac{w_4}{w_1} = 1 > 0 \,. 
\ena
Therefore, the DBI galileon theory automatically satisfies 
the constraints from GW170817 \cite{Baker:2017hug}
as well as  the stability conditions for the tensor perturbations.
.

\section{DOUBLE SCREENING MECHANISM} 
\label{screening}

Since the fifth force must be screened in the solar system scale, we will assume a flat static spherically symmetric metric, $ds^2 = -dt^2 + dr^2 + r^2 d \Omega^2$, to study screening mechanism. In addition, throughout this section, we shall assume $\phi = \phi(r)$. The equation of motion of the DBI galileon is given by
\bea
& &\nabla_{\mu} \left(\frac{1}{\sqrt{1 + 2 \lambda X}} \nabla^{\mu} \phi \right) \nn 
& & + \frac{1}{\Lambda^3} \nabla_{\mu} \left(\frac{1}{1 + 2 \lambda X} \left( \nabla^{\mu}\phi \square \phi - \nabla_{\nu}\phi\nabla^{\mu}\nabla^{\nu}\phi \right)\right) \nn
& &= \frac{g}{\mpl} \rho_m \,.
\ena
In this section we assume $a_2 \lambda = 1$ and $2 a_3 \lambda = 1/ \Lambda^3$  where $\Lambda$ is a constant with mass dimension.
Although $\lambda$ and $\Lambda$ can be chosen arbitrary values only
 for the screening mechanism, 
we have to set $\lambda=(\mpl^2 H_0^2)^{-1}$ and $\Lambda=(\mpl H_0^2)^{1/3}$
 if we also discuss the cosmological dynamics simultaneously.

\begin{center}
\begin{figure}[htb]
	\includegraphics[width=3in]{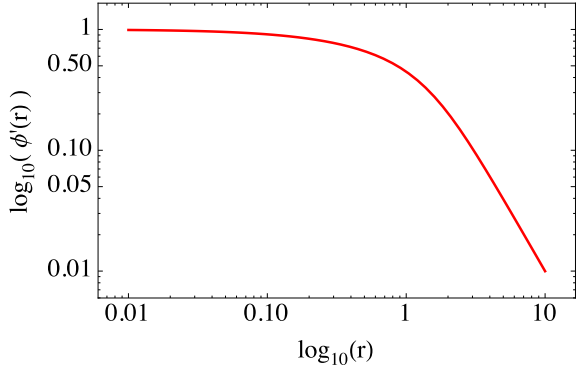} 
	\caption{A profile of the DBIonic galileon scalar field where we set $\lambda = \Lambda = \tilde{A} = 1$. The limit  $\lambda \phi^{\prime 2} \ll 1$ satisfies at large distances.} 
\label{profile}
\end{figure}
\end{center}

The fifth force, which arises from conformal interaction between matter and scalar field, in r-direction is given by \cite{Bloomfield:2014zfa,Fujii:2003pa,Burrage:2017qrf}
\bea
F^{r}_5 \equiv - \frac{A_{,\phi}}{A} \frac{d\phi}{dr} = - \frac{g}{\mpl} \frac{d\phi}{dr} \,. \label{fifthforce}
\ena
The equation of motion in a flat background metric becomes
\bea
& & \frac{1}{r^2} \frac{d}{dr} \left[r^3 \left(\frac{1}{\sqrt{1 - \lambda \phi^{\prime 2}}}
 \left(\frac{\phi^{\prime}}{r}\right) + \frac{2}{\Lambda^3}\frac{1}{1 - \lambda \phi^{\prime 2}} \left(\frac{\phi^{\prime}}{r}\right)^2 \right)\right] \nn 
 & & = \frac{g}{\mpl} \rho_m \,, \label{eom}
\ena
where $^{\prime} \equiv \frac{d}{dr}$. For a static point source i.e. $\rho_m = M \delta^{(3)} (\vec r)$, we obtain analytic solution of (\ref{eom}) as 
\bea
\phi^{\prime} (r) = \pm \sqrt{\frac{2 \tilde{A}^2}{\frac{4 \tilde{A} r}{\Lambda^3} + r^4 + 2 \tilde{A}^2 \lambda \pm \sqrt{r^8 + 8 \tilde{A} r^5 / \Lambda^3}}} \,, \label{sol1}
\ena
where $\tilde{A} \equiv g M/4 \pi \mpl$ which is proportional to a mass of the source. First of all, in order to obtain the real solution for the scalar field, $\lambda$ must be positive by considering the limit that $r \rightarrow 0$ (the solution becomes complex when $\lambda$ is negative). Thus, the spherically symmetric solution of $\phi(r)$ is admitted only in the DBIonic galileon. Secondly, for the outer $\pm$ sign, we choose a positive sign by considering the asymptotic form at large distances, namely there is only positive root when a linear term in the equation of motion dominates which is required in the screening mechanism (considering limits $\Lambda \rightarrow \infty$, $\lambda \phi^{\prime 2} \ll 1$ the linear term in the equation of motion dominates). Lastly, we also choose positive sign for the inner $\pm$ sign by considering the ratio of $F_{\phi}/ F_N \sim \phi^{\prime} r^2$, we find that for the negative sign the ratio can diverge, while the positive sign can provide the same order of magnitude between the fifth force and the Newtonian force at large distances. The solution (\ref{sol1}) with (both) positive signs and positive $\lambda$ is shown in Fig. \ref{profile}.

If $\Lambda \rightarrow \infty$ and $\lambda = \Lambda_{\ast}^{-4}$, the solution becomes
\bea
\phi^{\prime} (r) = \sqrt{\frac{\tilde{A}^2}{\lambda \tilde{A}^2 + r^4 }} = \frac{\Lambda_{\ast}^2}{\sqrt{1 + r^4/r^4_{\ast}}} \,. \label{dbionicsol}
\ena
This is the DBIonic solution \cite{Burrage:2014uwa} where a scale radius, $r_{\ast} \equiv \Lambda_{\ast}^{-1} \tilde{A}^{1/2}$, is analogous to the Vainshtein radius, and $\Lambda_{\ast}$ is a constant with a dimension of mass. 

If $\lambda \phi^{\prime 2} \ll 1$, the equation of motion (\ref{eom}) becomes the equation of motion of the galileon gravity \cite{Bloomfield:2014zfa}, where the solution is
\bea
\phi^{\prime} (r) = \frac{\Lambda^3 r}{4} \left(-1 + \sqrt{1 + \frac{r_V^3}{r^3}}\right) \,, \label{vainshteinsol}
\ena
and $r_V \equiv 2\tilde{A}^{1/3} \Lambda^{-1}$ is the Vainshtein radius. 

\begin{center}
\begin{figure}[htb]
	\includegraphics[width=3.5in]{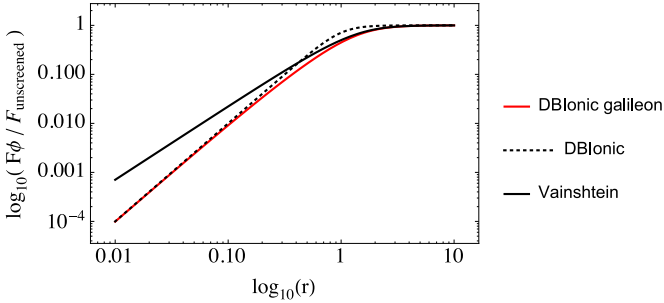} 
	\caption{The screening behaviour of the DBIonic galileon where we set $\lambda = \Lambda = \tilde{A} = 1$. The $F_{\rm unscreened} \equiv 2g^2 F_N$, and $F_N$ is the Newtonian force outside a point mass ($F_N = - M/8 \pi \mpl^2 r^2$).} 
\label{screeningplot}
\end{figure}
\end{center}

Therefore, the DBIonic galileon possesses (two) double screening mechanisms which has been firstly mentioned in \cite{Gratia:2016tgq}. By substituting the Eq. (\ref{sol1}) into the Eq. (\ref{fifthforce}), we find the fifth force of the DBIonic galileon model. A ratio of the fifth force and the Newtonian force or screening behaviour is shown in Fig. \ref{screeningplot}.

In Fig. \ref{screeningplot}, at large distances from the source, the ratio of the fifth force and the Newtonian force is close to the Vainshtein screening mechanism, whereas at small distances, the ratio approaches to the DBIonic screening. This characteristic behaviour can be understood by considering asymptotic limits of the DBIonic galileon solution. 

Applying the Vainshtein radius, $r_V =2\tilde{A}^{1/3} \Lambda^{-1}$, into the Eq. (\ref{sol1}), we find 
\bea
\phi^{\prime} (r) = \sqrt{\frac{\tilde{A}}{\tilde{A}\lambda + \frac{2r}{\Lambda^3}\left(1 + 2 \left(\frac{r}{r_V}\right)^3 + 2 \sqrt{\frac{r^3}{r_V^3}\left(\frac{r^3}{r_V^3} + 1\right)}\right)}} \,. \nn
\ena
At very large distances, $r \gg r_V$, the solution becomes
\bea
\phi^{\prime} (r) \simeq \frac{\tilde{A}}{r^2} \,.
\ena
This is proportional to inverse $r^2$, then the ratio between the fifth force and the Newtonian force is 
\bea
F_{\phi}/F_{N} \simeq 2 g^2 \,.
\ena
Thus, the fifth force is unscreened. 

According to the Fig. \ref{profile}, $\lambda \phi^{\prime 2} \ll 1$ at large distances, this is the limit to obtain the galileon gravity. Thus, the DBIonic galileon has the same behaviour as the Vainshtein mechanism. While at small distances $r \ll r_V$, we find
\bea
\phi^{\prime} (r) \simeq \sqrt{\frac{\tilde{A}}{\tilde{A} \lambda + \frac{2r}{\Lambda^3}}} \,.
\ena
If $2r/\Lambda^3 \gg \tilde A\lambda$, the solution becomes
\bea
\phi^{\prime} (r) \simeq \frac{2\tilde{A}}{r^{1/2} r_V^{3/2}} \,.
\ena
This is the same as the limit $r \ll r_V$ of the Vainshtein solution (\ref{vainshteinsol}), then the fifth force is screened in the same way as the Vainshtein screening:
\bea
F_{\phi}/F_{N} \simeq 4 g^2 \left(\frac{r}{r_V}\right)^{3/2} \ll 1 \,.
\ena
If $2r/\Lambda^3 \ll \tilde{A}\lambda$ or $r \rightarrow 0$, we find
\bea
\phi^{\prime} (r) \simeq \Lambda_{\ast}^2 \,.
\ena
The solution has the same asymptotic limit as the DBIonic solution (\ref{dbionicsol}), while the Vainshtein solution (\ref{vainshteinsol}) is diverge. This is the reason why we obtain the DBIonic screening at very small distances. In this case, the ratio of the fifth force and the Newtonian force is given by 
\bea
F_{\phi}/F_{N} \simeq 2g^2 \left(\frac{r}{r_{\ast}}\right)^2 ~ \ll 1 \,.
\ena
It is obvious to see that the fifth force is screened.

Therefore, a scale changing from the Vainshtein screening mechanism to DBIonic screening, $r_{sc}$, is defined as follows
\bea
\frac{2 r_{sc}}{\Lambda^3} &=& \tilde{A} \lambda \,, \nn
\therefore ~ r_{sc} &=& \frac{\Lambda^3}{2 \Lambda_{\ast}^4} \left( \frac{gM}{4\pi \mpl}\right) \,.
\ena
Consequently, this transition scale depends on a mass of the source \cite{Gratia:2016tgq}. For a small mass object, the $r_{sc}$ is small, then the fifth force from large distances approaches to the DBIonic screening slower, whereas for a massive object, it approaches to the DBIonic screening faster.

\section{Conclusions} 
\label{conclusions}

In this work we study cosmological dynamics of the DBI galileon theories up to cubic order without the scalar potential. The model has been divided into two regimes, one has positive signs in front of scalar field kinetic terms, so-called the DBI galileon, whereas negative signs lead to the DBIonic galileon. They are subclass of the Horndeski theory, where both models survive the GW170817 testing and the caustic singularities problem. By using the dynamical system approach we find that the DBIonic galileon does not have real fixed points, then there is no de Sitter solution, whereas the DBI galileon has two de Sitter fixed points (a) and (b)
when there is no matter coupling. 
Both fixed points are stable and are distinct  by initial conditions of the kinetic term, namely for negative $\dot\phi$ (scalar field value is decreasing with time) the solutions will converge to the fixed point (a) in which the kinetic energy vanishes at late-time, then it is similar to the cosmological constant. While for positive $\dot\phi$ the fixed point (b) is also an attractor, and the kinetic term still exists. This behaviour is confirmed by numerical simulations. We show that by choosing appropriated initial conditions the DBI galileon provides cosmological evolution from the radiation dominated era to the late-time accelerated expansion era correctly. We find necessary conditions that avoid the ghost and Laplacian instabilities in the model by considering cosmological perturbations. Both de Sitter fixed points yield the sound speed square greater than zero, thus they can avoid the Laplacian instability; however, the fixed point (b) cannot evade the ghost instability at late-time. Consequently, the viable cosmological solution of the DBI galileon model is only the fixed point (a).

When there is a coupling between a scalar field and matter field, we also find third fixed point for
the coupling constant with $g_{\rm cr}(\sim 1.788) \leq g<3$.
Although the fixed point shows a scaling behaviour such that $\Omega_m/\Omega_\phi$= 
non-zero constant, the solution has one unstable mode (one positive eigenvalue). As a result,
this fixed point may not be found for generic initial conditions.

In addition to the cosmological evolution, we analyze the screening mechanism of the DBI galileon and DBIonic galileon theories. We find that only DBIonic galileon has the spherically symmetric solution, while there is no real solution for the case of DBI galileon. The screening behaviour of the model has mixing of the DBIonic screening and the Vainshtein screening mechanism, which is called double screening. At large distances the fifth force has the same behaviour as the Vainshtein mechanism, whereas it becomes the DBIonic screening at very small distances from the source. Lastly, we find that the scale changing of the double screening depends on a mass of the source.

Although we explain the accelerating expansion of the universe by the DBI galileon theory 
and the screening of a scalar field by the DBIonic galileon theory, 
we could not find a model or theory which explains  both the  accelerating expansion 
as well as the screening simultaneously.
Based on the present work, investigation in a more general theory in which both cosmological evolution and screening mechanism are unified within the same framework is left for future work.

\section*{Acknowledgements}

S.P. (first author) and S.P. (second author) are supported by Rachadapisek Sompote Fund for Postdoctoral Fellowship, Chulalongkorn University. This work was supported by JSPS Grant-in-Aid for Scientific Research Grants No. 16K05362 and No. JP17H06359 (K. M.).

\begin{widetext}

\appendix

\section{Basic Equations of Horndeski Theory with conformal coupling} 
\label{basicHorn}

In this Appendix, we will show basic equations of the Horndeski theory with the conformal coupling between scalar field and matter field. The action is  
\bea
S = \int d^4 x \sqrt{-g}\left[\sum_{i = 2}^5 \mathcal{L}_i \right] + S_m (A^2 (\phi) g_{\mu\nu}, \psi_m)\,,
\ena
where
\beann
\mathcal{L}_2 &=& K(\phi,X) \,, \\
\mathcal{L}_3 &=& - G_3 (\phi,X) \square \phi \,, \\
\mathcal{L}_4 &=& G_4 (\phi,X) R + G_{4X} \nabla^{\mu}_{[\mu} \phi \nabla^{\nu}_{\nu]} \phi\,, \\
\mathcal{L}_5 &=& G_5(\phi,X) G_{\mu\nu} \nabla^{\mu} \nabla^{\nu} \phi - \frac{1}{6} G_{5X} \nabla^{\mu}_{[\mu} \phi \nabla^{\nu}_{\nu} \phi \nabla^{\rho}_{\rho]}\,.
\enann
The antisymmetrisation does not include the factor $1/n! $, and $X = - \frac{1}{2} (\nabla \phi)^2$. We use short-hand notation as $\phi_{\mu} \equiv \nabla_{\mu} \phi$, $\phi_{\mu\nu} \equiv \nabla_{\mu}\nabla_{\nu} \phi$, and $K_X \equiv \partial K / \partial X$. Variation of the above action with respect to $g^{\mu\nu}$ and $\phi$ yield field equation and equation of motion of scalar field, respectively: 
\beann
\delta S = \int d^4 x \sqrt{-g} \left( \varepsilon_{\mu\nu} \delta g^{\mu\nu} + \varepsilon_{\phi} \delta \phi \right) + \delta S_m \,,
\enann
the field equation is 
\bea
\sum_{i = 2}^5 \varepsilon^i_{\mu\nu} = \frac{1}{2}T^m_{\mu\nu} \,,
\ena
we define 
\bea
\varepsilon^i_{\mu\nu} = \frac{1}{2} (E^i_{\mu\nu} + E^i_{\nu\mu}) \,,
\ena
where
\beann
 E^2_{\mu\nu} &=& - \frac{1}{2} (Kg_{\mu\nu} + K_X \phi_{\mu} \phi_{\nu}) \,, \\
 E^3_{\mu\nu} &=& \frac{1}{2}(G_{3X} \square \phi \phi_{\mu} \phi_{\nu} + 2 G_{3;\mu} \phi_{\nu} - g_{\mu\nu} G_{3;\rho} \phi^{\rho}) \,, \\
E^4_{\mu\nu} &=& - \frac{1}{2}\left(-2G_4 G_{\mu\nu} + G_{4X} R \phi_{\mu} \phi_{\nu} + g_{\mu\nu} G_{4X} \phi^{[\rho}_{\rho} \phi^{\sigma]}_{\sigma} + G_{4XX} \phi_{\mu} \phi_{\nu} \phi^{[\rho}_{\rho} \phi^{\sigma]}_{\sigma}\right) + (G_{4\phi} \phi^{\alpha})_{;[\alpha} g_{\mu]\nu} \\ 
& & - (G_{4X} \phi_{\beta})_{;[\alpha}g_{\mu]\nu} \phi^{\alpha\beta} - G_{4X} \phi^{\alpha} g_{\mu[\nu} R_{\beta]\rho ~ \alpha}^{~~~\beta} \phi^{\rho} + (G_{4X} \phi_{\alpha})^{;\alpha} \phi^{\beta}_{[\beta} g_{\mu]\nu} - 2G_{4X;[\mu} \phi^{\alpha}_{\alpha]} \phi_{\nu} + 2G_{4X} \phi_{\nu} R_{\mu \alpha}\phi^{\alpha} \,, \\
E^5_{\mu\nu} &=& - \frac{1}{2} \left[-\frac{1}{2}g_{\mu\nu} G_5 R \square \phi + G_{5X} \phi_{\mu} \phi_{\nu} G_{\alpha\beta} \phi^{\alpha\beta} + g_{\mu[\nu} \phi^{\beta}_{\beta} \nabla_{\alpha]}(G_{5\phi} \phi^{\alpha}) - g_{\mu[\nu} \phi^{\beta}_{\beta} \nabla_{\alpha]} (G_{5X} \phi_{\rho}) \phi^{\rho\alpha} + 2G_5 R_{\mu \beta} \phi _{\nu}^{\beta} \right. \\ 
& & + 2 G_{5;\mu} R_{\nu\beta} \phi^{\beta} - 2G_5^{;\alpha} R_{\alpha\mu\nu \beta} \phi^{\beta} - 2 G_{5;\alpha} R^{\alpha\beta} \phi_{\beta} g_{\mu\nu} - 4 G_5 G_{\mu \alpha } \phi_{\nu}^{\alpha} + 2 (G_5 \phi_{\mu})_{;\alpha} G_{\nu}^{\alpha} - (G_5 \phi^{\alpha})_{;\alpha} G_{\mu\nu} \\
& & + G_5 (R_{\mu\nu} \square \phi - R \phi_{\mu\nu}) - \frac{1}{6} g_{\mu\nu} G_{5X} \phi^{[\alpha}_{\alpha} \phi^{\beta}_{\beta} \phi^{\rho]}_{\rho} - \frac{1}{6} G_{5XX} \phi_{\mu} \phi_{\nu} \phi^{[\alpha}_{\alpha} \phi^{\beta}_{\beta} \phi^{\rho]}_{\rho} + \frac{1}{2} G_{5X;\alpha} \phi^{\alpha} g_{\mu [ \nu} \phi^{\rho}_{\rho} \phi^{\sigma}_{\sigma]} \\ 
& & - G_{5X;[\mu} \phi^{\rho}_{\rho} \phi^{\sigma}_{\sigma]} \phi_{\nu}
\left.+ \frac{1}{2}G_{5X} \square \phi g_{\mu [ \nu} \phi^{\rho}_{\rho} \phi^{\sigma}_{\sigma]} - G_{5X} \phi_{\mu} \phi^{\alpha}_{[\alpha} R_{\nu \beta] ~ \rho}^{~~~ \beta} \phi^{\rho} - G_{5X} g_{\mu [\nu} \phi^{\alpha}_{\alpha} R_{\beta]\rho ~ \sigma}^{~~~ \beta} \phi^{\rho} \phi^{\sigma} \right] \,,
\enann
and $T^m_{\mu\nu}$ is the energy-momentum tensor of (non-relativistic) matter and radiation. The equation of motion of scalar field is
\bea
\sum_{i = 2}^5 \varepsilon^i_{\phi} = - \frac{A_{,\phi}}{A} T^m \,,
\ena
where
\beann
\varepsilon^2_{\phi} &=& K_{\phi} + (K_X \phi^{\mu})_{;\mu} \,, \\
\varepsilon^3_{\phi} &=& -G_{3\phi} \square \phi - \square \phi (G_{3X}\phi^{\mu})_{;\mu} + G_{3X} R_{\mu\nu} \phi^{\mu} \phi^{\nu} - (G_{3\phi} \phi_{\mu})^{;\mu} + (G_{3X} \phi^{\nu})^{;\mu} \phi_{\mu\nu} \,, \\
\varepsilon^4_{\phi} &=& G_{4\phi} R + (G_{4X}\phi_{\nu})^{;\nu} R + G_{4X\phi} \phi^{[\mu}_{\mu} \phi^{\nu]}_{\nu} + (G_{4XX}\phi_{\rho})^{;\rho} \phi^{[\mu}_{\mu} \phi^{\nu]}_{\nu} - 2 G_{4XX} \phi^{[\mu}_{\mu} R^{\beta]}_{~~\alpha \beta \nu} \phi^{\alpha} \phi^{\nu} \\
& & + 2( G_{4X\phi} \phi_{\mu})^{;[\mu} \phi^{\nu]}_{\nu} - 2 (G_{4XX} \phi_{\alpha})^{;[\nu}\phi^{\mu]}_{\mu} \phi^{\alpha}_{\nu} - 4 G_{4X}^{;\nu} R_{\mu\nu}  \phi^{\mu} - 2 G_{4X} R_{\mu\nu} \phi^{\mu\nu} \,, \\
\varepsilon^5_{\phi} &=& G_{5\phi} G_{\mu\nu} \phi^{\mu\nu} + (G_{5X} \phi_{\alpha})^{;\alpha} G_{\mu\nu}\phi^{\mu\nu} + G_{\mu\nu} (G_{5\phi} \phi^{\mu})^{;\nu} - G_{\mu\nu} (G_{5X} \phi_{\alpha})^{;\nu} \phi^{\mu\alpha} + G_{5X} R_{\alpha\beta\mu\nu} G^{\alpha\nu}\phi^{\beta}\phi^{\mu} \\
& & - \frac{1}{6} G_{5X\phi} \phi^{[\mu}_{\mu} \phi^{\nu}_{\nu} \phi^{\rho]}_{\rho} - \frac{1}{6} (G_{5XX} \phi^{\alpha})_{;\alpha} \phi^{[\mu}_{\mu} \phi^{\nu}_{\nu} \phi^{\rho]}_{\rho} + \frac{1}{2}G_{5XX} \phi^{\mu} \phi^{\nu} \phi^{[\alpha}_{\alpha} \phi^{\beta}_{\beta} R^{\rho]}_{~~\mu \rho \nu} - G_{5X;\alpha} \phi^{[\beta}_{\beta} R^{\alpha\mu]}_{~~~ \mu \nu} \phi^{\nu} \\
& & - \frac{1}{2} (G_{5X\phi} \phi_{\mu})^{;[\mu} \phi^{\nu}_{\nu} \phi^{\rho]}_{\rho} + \frac{1}{2} (G_{5XX} \phi^{\sigma})^{;[\alpha} \phi^{\beta}_{\beta} \phi^{\rho]}_{\rho} \phi_{\alpha\sigma} - \frac{1}{2} G_{5X} R^{[\alpha\beta}_{~~~ \beta \sigma} R_{\alpha~~ \rho \lambda}^{~ \rho]} \phi^{\sigma}\phi^{\lambda} - G_{5X} \phi^{[\alpha}_{\sigma} \phi^{\beta}_{\beta} R_{\alpha ~\rho}^{~ \rho] ~ \sigma} \,.
\enann
To study cosmological evolution, we assume that the scalar field is homogenous, i.e. $\phi = \phi(t)$, and using the flat FLRW metric, $ds^2 = - dt^2 + a^2 (t)d \bf{x}^2$. Since the energy-momentum tensor is
\bea
T^H_{\mu\nu} = -\frac{2}{\sqrt{-g}}\frac{\delta S_{H}}{\delta g^{\mu\nu}} = -2 \sum_{i = 2}^5 \varepsilon^i_{\mu\nu} \,,
\ena
the energy density and pressure of the Horndeski theory are as follows
\bea
\rho_{H} &=& 2X K_X - K + 6H \dot \phi X G_{3X} - 2 X G_{3\phi} - 6H^2 G_4 - 6 H \dot\phi G_{4\phi} + 24 H^2 X G_{4X} + 24 H^2 X^2 G_{4XX} \nn
& & - 12 H \dot\phi X G_{4X\phi} + 10 H^3 X \dot\phi G_{5X} - 18 H^2 X G_{5\phi} -12 H^2 X^2 G_{5X\phi} + 4 H^3 X^2 \dot\phi G_{5XX} \,. \\
P_{H} &=& K - 2X G_{3\phi} - 2X \ddot\phi G_{3X} + G_4 (4 \dot H + 6H^2) + G_{4X} (-12 H^2 X - 4 H \dot\phi \ddot\phi - 8 \dot H X) + G_{4\phi} (2 \ddot\phi + 4 H \dot\phi) \nn
& & + G_{4XX} (-8 H X \dot\phi \ddot\phi) + G_{4X\phi} (4 X \ddot\phi - 8 H X \dot\phi) + G_{4\phi\phi} (4X)+ G_{5X} ( - 6 H^2 X \ddot \phi - 4 H \dot H \dot\phi X - 4 H^3 X \dot\phi) \nn
& & + G_{5\phi} (4H\dot\phi \ddot\phi + 4 \dot H X + 6 H^2 X) + G_{5XX} (-4H^2 X^2 \ddot\phi) + G_{5\phi\phi} (4 H X \dot \phi) + G_{5\phi X} (4 H X \dot\phi \ddot\phi - 4 H^2 X^2) \,, \nn
\ena
where $X = \dot\phi^2 / 2$. The equation of motion of scalar field under the flat FLRW metric becomes
\bea
\varepsilon^2_{\phi} &=& K_{\phi} - 2 X K_{X\phi} - 2X \ddot \phi K_{XX} - (\ddot\phi + 3 H \dot\phi) K_X \,, \\
\varepsilon^3_{\phi} &=& G_{3\phi} (2\ddot\phi + 6H \dot\phi) + G_{3X} (- 6 H \dot\phi \ddot\phi - 18 H^2 X - 6 \dot H X) + G_{3XX} (- 6 H X \dot\phi \ddot\phi) + G_{3\phi\phi} (2X) \nn
& & + G_{3X\phi} ( - 6 H X \dot\phi + 2 X \ddot\phi) \,, \\
\varepsilon^4_{\phi} &=& G_{4\phi} (6 \dot H + 12 H^2) + G_{4X} (-6 H^2 \ddot \phi - 12 H \dot H \dot\phi -18 H^3 \dot\phi) + G_{4XX} (-24 H \dot H X \dot\phi - 48 H^2 X \ddot\phi -36 H^3 X \dot\phi) \nn
& & + G_{4X\phi} (12 \dot H X + 18 H \dot\phi \ddot\phi + 36 H^2 X) + G_{4XXX} (-24 H^2 X^2 \ddot \phi) + G_{4XX\phi} ( - 24 H^2 X^2 + 12 H X \dot\phi \ddot\phi) \nn
& & + G_{4X\phi\phi} (12 H X \dot\phi) \,, \\
\varepsilon^5_{\phi} &=& G_{5\phi} (6 H^2 \ddot\phi + 12 H \dot H \dot\phi + 18 H^3 \dot\phi) + G_{5X} (-18 H^2 \dot H X - 18 H^4 X - 6 H^3 \dot\phi \ddot\phi) \nn
& & + G_{5XX} (-14 H^3 X \dot\phi \ddot\phi - 12 H^4 X^2 - 12 H^2 \dot H X^2 ) + G_{5X\phi} (12 H \dot H X \dot\phi + 14 H^3 X \dot\phi + 30 H^2 X \ddot\phi) + G_{5\phi\phi} (6 H^2 X) \nn
& & + G_{5XXX} (-4 H^3 X^2 \dot\phi \ddot\phi) + G_{5XX\phi} (12 H^2 X^2 \ddot\phi - 4 H^3 X^2 \dot\phi) + G_{5X\phi\phi} (12 H^2 X^2) \,.
\ena
Calculations in this Appendix agree with the results in \cite{Gomes:2015dhl,Padilla:2012ze}. Nevertheless, after the discovery of GW170817 the higher order terms have been constrained. The viable Horndeski theory is allowed up to cubic order ($\mathcal{L}_3$) with the conformal coupling to gravity ($G_4(\phi) R$) \cite{Baker:2017hug,Kase:2018iwp}.

\section{Components of the matrix $\mathcal{M}$}
\label{matrixperturbations}

For the autonomous equations (\ref{autoeq1}) - (\ref{autoeq3}), we find components of the matrix $\mathcal{M}$ as follows
\bea
\frac{\partial F}{\partial x_1} &=& -\frac{1}{\lambda  x_2 \sqrt{\lambda  x_1^2+1} \xi^2} \left(2 g \left[-3 \lambda ^2 x_1^7+6 \lambda  x_1^2 x_2 \left(12 \lambda  x_2^2 (x_3-1) \sqrt{\lambda  x_1^2+1}+1\right) \right.\right.\nn
& & \left.\left.+2 \lambda  x_1 \left(2 \sqrt{\lambda  x_1^2+1}-9 \lambda  x_2^2 (x_3-1)\right)+12 x_2 \left(1-3 \lambda  x_2^2 (x_3-1) \sqrt{\lambda  x_1^2+1}\right) \right.\right. \nn
& & \left.\left. +9 \lambda ^2 x_1^8 x_2 \sqrt{\lambda  x_1^2+1}-3 \lambda  x_1^5 \left(6 \lambda  x_2^2 \left(\lambda ^2 (x_3-1)-2 (x_3-5) \sqrt{\lambda  x_1^2+1}\right)+5\right) \right.\right.\nn
& & \left.\left. +x_1^6 x_2 \left(24 \lambda ^3-9 \lambda  \sqrt{\lambda  x_1^2+1}\right)+18 \lambda ^2 x_1^4 x_2 \left(6 \lambda  x_2^2 (x_3-1) \sqrt{\lambda  x_1^2+1}+1\right)\right.\right.\nn
& & \left.\left.+4 x_1^3 \left(\lambda ^2 \sqrt{\lambda  x_1^2+1}+9 \lambda  x_2^2 \left((x_3-3) \sqrt{\lambda  x_1^2+1}-\lambda ^2 (x_3-1)\right) -3\right)\right] \right. \nn
& & \left. +3 \left(3 \lambda ^2 x_1^9-3 \lambda  x_1^7-66 \lambda ^2 x_1^6 x_2-4 x_1^2 x_2 \left(-4 \lambda ^2 \sqrt{\lambda  x_1^2+1}+3 \lambda  x_2^2 (x_3-3) \sqrt{\lambda  x_1^2+1}+3\right) \right.\right.\nn
& & \left.\left.+4 x_1 \left(\sqrt{\lambda  x_1^2+1}+\lambda  x_2^2 (x_3-3)\right)+4 \lambda  x_2 \sqrt{\lambda  x_1^2+1}+2 x_1^5 \left(4 \lambda ^2 \sqrt{\lambda  x_1^2+1}+x_2^2 \left(3 \lambda ^3 (x_3-15) \right.\right.\right.\right.\nn 
& & \left.\left.\left.\left.-3 \lambda  (x_3-3) \sqrt{\lambda  x_1^2+1}\right)-3\right)-6 \lambda  x_1^4 x_2 \left(-2 \lambda ^2 \sqrt{\lambda  x_1^2+1}+6 \lambda  x_2^2 (x_3-3) \sqrt{\lambda  x_1^2+1}+13\right) \right.\right.\nn
& & \left.\left.+2 \lambda  x_1^3 \left(6 \sqrt{\lambda  x_1^2+1}+\lambda  x_2^2 (5 x_3-51)\right)\right)\right) \,, \\
\frac{\partial F}{\partial x_2} &=& \frac{\left(\lambda  x_1^2+1\right)}{\lambda  x_2^2 \xi^2} \left(2 g \left[9 \lambda ^2 x_1^6 x_2^2 (x_3-1)+2 \lambda  x_1^2 \left(3 \lambda  x_2^2 (x_3-1) \sqrt{\lambda  x_1^2+1}+1\right) \right.\right.\nn
& & \left.\left. +6 \lambda  x_2^2 (x_3-1) \sqrt{\lambda  x_1^2+1} -24 x_1 x_2 \sqrt{\lambda  x_1^2+1}+3 x_1^4 \left(\sqrt{\lambda  x_1^2+1}+3 \lambda  x_2^2 (x_3-5)\right)+2\right] \right. \nn
& & \left.+3 x_1^2 \left(2 \lambda  x_1^2 -2 \left(\lambda  x_2^2 (x_3+9) \sqrt{\lambda  x_1^2+1}-1\right) -24 x_1 x_2 \sqrt{\lambda  x_1^2+1} \right.\right.\nn
& & \left.\left.+3 x_1^4 \left(\sqrt{\lambda  x_1^2+1}-\lambda  x_2^2 (x_3-3)\right)\right)\right) \,, \\
\frac{\partial F}{\partial x_3} &=&\frac{3 x_2 \left(\lambda  x_1^2+1\right) \left(2 g \left(\lambda  x_1^2+1\right)-x_1^2\right)}{\xi} \,, \\
\frac{\partial G}{\partial x_1} &=& \frac{x_1}{\lambda  x_2 \sqrt{\lambda  x_1^2+1} \xi^2} \left(g \left[-3 \lambda  x_1^6+24 \lambda ^2 x_1^5 x_2+2 \lambda  x_1^2 \left(2 \sqrt{\lambda  x_1^2+1}-15 \lambda  x_2^2 (x_3-1)\right) \right.\right. \nn
& & \left.\left.+12 x_1 x_2 \left(3 \lambda  x_2^2 (x_3-1) \sqrt{\lambda  x_1^2+1}-1\right)+4 \left(\sqrt{\lambda  x_1^2+1}-3 \lambda  x_2^2 (x_3-1)\right)+9 \lambda  x_1^7 x_2 \sqrt{\lambda  x_1^2+1} \right.\right.\nn
& & \left.\left.+6 x_1^4 \left(3 \lambda  x_2^2 \left((x_3-9) \sqrt{\lambda  x_1^2+1}-\lambda ^2 (x_3-1)\right)-1\right)+6 \lambda  x_1^3 x_2 \left(18 \lambda  x_2^2 (x_3-1) \sqrt{\lambda  x_1^2+1}+1\right)\right] \right.\nn
& & \left.-2 \lambda -45 \lambda  x_1^5 x_2+2 x_1^2 \left(-\lambda ^2+6 \sqrt{\lambda  x_1^2+1}+6 \lambda  x_2^2 (x_3-15)\right)+42 \lambda  x_1 x_2 \sqrt{\lambda  x_1^2+1} \right.\nn
& & \left.+3 \lambda  x_1^4 \left(2 \sqrt{\lambda  x_1^2+1}+3 \lambda  x_2^2 (x_3-15)\right)-18 x_1^3 x_2 \left(-\lambda ^2 \sqrt{\lambda  x_1^2+1}+3 \lambda  x_2^2 (x_3-3) \sqrt{\lambda  x_1^2+1}+3\right)\right) , \\
\frac{\partial G}{\partial x_2} &=& -\frac{1}{\lambda  x_2^2 \xi^2} \left(g x_1^2 \left[9 \lambda ^2 x_1^6 x_2^2 (x_3-1)+2 \lambda  x_1^2 \left(3 \lambda  x_2^2 (x_3-1) \sqrt{\lambda  x_1^2+1}+1\right) \right.\right.\nn
& & \left.\left.+6 \lambda  x_2^2 (x_3-1) \sqrt{\lambda  x_1^2+1}-24 x_1 x_2 \sqrt{\lambda  x_1^2+1}+3 x_1^4 \left(\sqrt{\lambda  x_1^2+1}+3 \lambda  x_2^2 (x_3-5)\right)+2\right]+27 \lambda  x_1^8 x_2^2 \right.\nn
& & \left.-3 \lambda  x_1^6-36 \lambda  x_1^5 x_2^3 (x_3+3)+24 \lambda  x_1^3 x_2-2 \sqrt{\lambda  x_1^2+1}+2 x_1^2 \left(\lambda +36 x_2^2\right) \left(\lambda  x_2^2 (x_3+3)-\sqrt{\lambda  x_1^2+1}\right) \right.\nn
& & \left.-24 x_1 x_2 \left(\lambda  x_2^2 (x_3+3) \sqrt{\lambda  x_1^2+1}-1\right)+3 x_1^4 \left(\lambda  x_2^2 (x_3-3) \sqrt{\lambda  x_1^2+1}-1\right)+6 \lambda  x_2^2+2 \lambda  x_2^2 x_3\right) \,, \\
\frac{\partial G}{\partial x_3} &=& -\frac{x_2 \left(3 g \left(\lambda  x_1^4+x_1^2\right)+\sqrt{\lambda  x_1^2+1}-6 x_1 x_2\right)}{\xi} \,, \\
\frac{\partial \mathcal{H}}{\partial x_1} &=& -\frac{2 x_1 x_3 }{\lambda  x_2^2 \sqrt{\lambda  x_1^2+1} \xi^2} \left(g \left[-3 \lambda  x_1^6+24 \lambda ^2 x_1^5 x_2+2 \lambda  x_1^2 \left(2 \sqrt{\lambda  x_1^2+1}-15 \lambda  x_2^2 (x_3-1)\right) \right.\right.\nn 
& & \left.\left.+12 x_1 x_2 \left(3 \lambda  x_2^2 (x_3-1) \sqrt{\lambda  x_1^2+1}-1\right)+4 \left(\sqrt{\lambda  x_1^2+1}-3 \lambda  x_2^2 (x_3-1)\right)+9 \lambda  x_1^7 x_2 \sqrt{\lambda  x_1^2+1} \right.\right.\nn
& & \left.\left.+6 x_1^4 \left(3 \lambda  x_2^2 \left((x_3-9) \sqrt{\lambda  x_1^2+1}-\lambda ^2 (x_3-1)\right)-1\right)+6 \lambda  x_1^3 x_2 \left(18 \lambda  x_2^2 (x_3-1) \sqrt{\lambda  x_1^2+1}+1\right)\right]\right. \nn 
& & \left.-2 \lambda -45 \lambda  x_1^5 x_2+2 x_1^2 \left(-\lambda ^2+6 \sqrt{\lambda  x_1^2+1}+6 \lambda  x_2^2 (x_3-15)\right)+42 \lambda  x_1 x_2 \sqrt{\lambda  x_1^2+1} \right.\nn
& & \left.+3 \lambda  x_1^4 \left(2 \sqrt{\lambda  x_1^2+1}+3 \lambda  x_2^2 (x_3-15)\right)-18 x_1^3 x_2 \left(-\lambda ^2 \sqrt{\lambda  x_1^2+1}+3 \lambda  x_2^2 (x_3-3) \sqrt{\lambda  x_1^2+1}+3\right)\right) , \\
\frac{\partial \mathcal{H}}{\partial x_2} &=& \frac{2 x_3 }{\lambda  x_2^3 \xi^2} \left(g x_1^2 \left[9 \lambda  x_1^7 x_2+4 \lambda  x_1^2-36 x_1 x_2 \left(\sqrt{\lambda  x_1^2+1}+x_2^2 (\lambda -\lambda  x_3)\right) \right.\right. \nn
& & \left.\left. +6 x_1^4 \left(\sqrt{\lambda  x_1^2+1}-12 \lambda  x_2^2\right) +6 \lambda  x_1^3 x_2 \left(\sqrt{\lambda  x_1^2+1}+6 \lambda  x_2^2 (x_3-1)\right)+4\right]-6 \lambda  x_1^6 \right.\nn
& & \left. +54 \lambda  x_1^3 x_2 -4 \sqrt{\lambda  x_1^2+1} -4 x_1^2 \sqrt{\lambda  x_1^2+1} \left(\lambda +36 x_2^2\right) +9 \lambda  x_1^7 x_2 \sqrt{\lambda  x_1^2+1} \right.\nn
& & \left.+6 x_1^5 x_2 \left(\lambda ^2+3 \sqrt{\lambda  x_1^2+1}-3 \lambda  x_2^2 (x_3-3)\right) -6 x_1^4 \left(12 \lambda  x_2^2 \sqrt{\lambda  x_1^2+1}+1\right)+48 x_1 x_2\right) \,, \\
\frac{\partial \mathcal{H}}{\partial x_3} &=& - \frac{2}{\lambda  x_2^2 \xi}\left(g x_1^2 \left[3 \lambda  x_1^3 x_2-3 \lambda ^2 x_1^2 x_2^2 (2 x_3-1)+\sqrt{\lambda  x_1^2+1}+3 \lambda  x_2^2 (1-2 x_3)\right] \right. \nn & & \left. + \left(-3 \lambda  x_1^4 x_2^2-\lambda  x_1^2+6 x_1 x_2 \left(\sqrt{\lambda  x_1^2+1}+\lambda  x_2^2 (2 x_3-1)\right)+\lambda  x_2^2 (1-2 x_3) \sqrt{\lambda  x_1^2+1} \right.\right. \nn & & \left.\left. +3 \lambda  x_1^3 x_2 \sqrt{\lambda  x_1^2+1}-1\right)\right) \,,
\ena
where
\bea
\xi \equiv 3x_1 (x_1^3 - 4x_2) + 2 \sqrt{\lambda x_1^2 + 1} \,.
\ena

\end{widetext}


\end{document}